\documentclass[aps,pre,twocolumn,showpacs,floatfix]{revtex4-1}
\usepackage{latexsym}
\usepackage{graphicx,color}
\usepackage{hyperref}

\begin{document}
\title{Effective merging dynamics of two and three fluid vortices:\\
Application to two-dimensional decaying turbulence}

\author{Cl\'ement Sire, Pierre-Henri Chavanis, and Julien
Sopik} \email{clement.sire@irsamc.ups-tlse.fr\,;
chavanis@irsamc.ups-tlse.fr}
%\affiliation{Universit\'e de Toulouse; UPS; \\Laboratoire de
%Physique Th\'eorique (IRSAMC); F-31062 Toulouse, France}
%\affiliation{CNRS; LPT (IRSAMC); F-31062 Toulouse, France}
\affiliation{{Laboratoire de Physique Th\'eorique -- IRSAMC,
Universit\'e de Toulouse (UPS) and CNRS, F-31062 Toulouse, France}}
%\date{\today}
\begin{abstract}
We present a kinetic theory of two-dimensional decaying turbulence
in the context of two-body and three-body vortex merging processes.
By introducing the equations of motion for two or three vortices in
the effective noise due to all the other vortices, we demonstrate
analytically that a two-body mechanism becomes inefficient at low
vortex density $n\ll 1$. When the more efficient three-body vortex
mergings are considered {(involving vortices of different signs)},
we show that $n\sim t^{-\xi}$, with $\xi=1$. We generalize this
argument to three-dimensional geostrophic turbulence, finding
$\xi=5/4$, in excellent agreement with direct Navier-Stokes
simulations [J.\,C.~McWilliams \emph{et al.}, J. Fluid Mech. {\bf
401}, 1 (1999)].
\end{abstract}
\maketitle

\section{Introduction and background}\label{sec1turb}
Two-dimensional {freely} decaying turbulence (\textit{i.e.} when an
external forcing has been switched off at the initial time $t=0$)
exhibits fascinating properties \cite{Tabeling2002}: starting from
an initial incoherent turbulent background, robust coherent
structures (the vortices) soon emerge
\cite{fornberg,blsb,McWilliams1984,Benzi1988,McWilliams1990}.
Different types of structures can be observed: {monopolar} vortices
rotating in either direction \cite{Brown1974,Aref1980,Couder1983},
quasi ballistic dipoles (a vortex/antivortex pair)
\cite{Couder1983,Legras1988} and even rare tripoles (a pair of two
like-sign vortices separated by an antivortex) \cite{Legras1988}.
These vortices interact due to their mutual advection which amounts
to a long-range interaction. {The vortices are characterized by
their radius $a(t)$ and by their core (or peak) vorticity
$\omega(t)$ while their precise vorticity profile is not
particularly relevant in the process of 2D decaying turbulence.}
When two like-sign vortices approach at a critical distance of {the
order of} their radius $a(t)$, a complex merging process occurs
\cite{Melander1988} and results in the formation of a larger vortex.
As a consequence of the merging processes, the vortex density $n(t)$
decays with time while their typical size $a(t)$ increases.

Numerical simulations of the Navier-Stokes (NS) equation
\cite{McWilliams1990,Carnevale1991,Weiss1993,Bracco2000,cn,Laval2001,bokhoven,dritschel}
and of the point vortex model presented below
\cite{Benzi1992,Weiss1993,Sire2000} are consistent with experiments
\cite{Tabeling1991,Cardoso1994,Hansen1998,clercx} in finding a
power-law decay of the vortex density $n\sim t^{-\xi}$, with $\xi$
in the range $0.6-1.0$, although it is most often found around
$\xi\sim 0.70-0.75$. For an inviscid flow, the dimensional Batchelor
argument \cite{Batchelor1969} which assumes that the energy per unit
area, $E\sim (1/L^2) \int {\bf u}^2 \,d^2{\bf x}\sim \omega^2 n
a^4$, is the sole invariant predicts $n\sim 1/(Et^2)$. This implies
$\xi=2$, in clear disagreement with numerical and experimental data.
However, in \cite{Carnevale1991}, and in agreement with subsequent
theoretical and experimental work, the average peak vorticity
$\omega$ was also found to remain almost constant during the
dynamics {(see Appendix \ref{secAturb})}.

The theoretical determination of the exponent $\xi$ has been a
subject of active research and many theories, based on different
physical arguments -- and leading in general to different results --
have been proposed over the years
\cite{clem,dritschel,ifo,Trizac1998,Pomeau1996,Sire2000,yakhot}.
Recently, \cite{lacasce2008} reviewed several theories and showed
that some of them are inconsistent (except that of \cite{Pomeau1996}
and \cite{Sire2000}; \cite{dritschel} was not reviewed, since not yet
published at the time) with the fact that the vortex typical velocity
$v\sim \omega\, n^{1/2} a^2\sim \sqrt E$ is constant
\cite{Sire2000}.

The vortex density satisfies a general equation of the form
\begin{equation}
dn/dt=-n/\tau_{m},
\end{equation}
where the merging time $\tau_{m}(t)$ is, by definition, the typical
time needed for two vortices to actually merge. Since we expect a
power-law behavior for $n(t)\sim t^{-\xi}$, we conclude that
$\tau_{m} \sim t$. Note that checking that $\tau_{m} \sim t$ in NS
simulations or experiments is a good test {to know} whether the
system has entered the asymptotic scaling regime \cite{Sire2000}. In
units of the collision time,
\begin{equation}
\tau \sim R/v\sim (\omega\, n a^2)^{-1},
\end{equation}
which is the typical time needed for a vortex of velocity $v\sim
\omega\, n^{1/2} a^2$ \cite{Sire2000} to travel the average inter
vortex distance $R\sim n^{-1/2}$, we obtain the most general form
\begin{equation}
\label{merg} \tau_{m} \sim \frac{\tau}{\left(n a^2\right)^\alpha}\sim
\tau\left(\frac{R}{a}\right)^{2\alpha}\sim t,
\end{equation}
where the exponent $\alpha$ cannot be determined on purely
dimensional grounds since $n a^2$ or $R/a$ are dimensionless. The
area fraction covered by the vortices {$na^2$} is actually the only
available dimensionless quantity {\cite{Sire2000}}. Now imposing
energy and peak vorticity conservation, Eq.~(\ref{merg}) leads to
\begin{equation}
\xi=\frac{2}{1+\alpha},
\end{equation}
so that determining $\xi$ amounts to determining $\alpha$, or the
merging time $\tau_{m}$. Note that Batchelor's argument
\cite{Batchelor1969} assumes that $\tau_{m} \sim \tau\sim t$,
\emph{i.e.} $\alpha=0$, after which the energy conservation and
Eq.~(\ref{merg}) leads to $\xi=2$, and $a\sim R\sim t$, in clear
disagreement with numerical and experimental data. Hence, the
necessity of a more general scaling theory arises from the fact that
the two length scales $a$ (typical vortex radius) and $R$ (typical
inter-vortex distance) do not scale in the same way with time. {In
other words}, the mean free time $\tau$ and the merging time
$\tau_m$ are not equivalent quantities, especially at low vortex
density.

We now consider the effective vortex model introduced in
\cite{Carnevale1991}. The exact Hamiltonian for point-like vortices
can be derived from the inviscid NS equation \cite{Kirchhoff1877}
and is reminiscent of two-dimensional electrostatics
\begin{equation}\label{hamilt}
{\cal H} = - \sum_{i \ne j}\gamma_i \gamma_j \ln(|{\bf r}_i-{\bf r}_j|),
\end{equation}
where $\gamma_i\sim \pm\omega a^2$ are the vortex circulations. The
crucial difference with electrostatics resides in the fact that the
vortex coordinates $(x_i,y_i)$ are the conjugate variables, so that
the resulting equations of motion are
\begin{eqnarray}
\label{eqx} \frac{d x_i}{dt} &=& \gamma_i^{-1}\frac{\partial H}{\partial y_i}
= - \sum_{i \ne
j}\gamma_j \frac{y_i-y_j}{r_{ij}^2},
\\
\label{eqy} \frac{d y_i}{dt} &=& - \gamma_i^{-1}\frac{\partial
H}{\partial x_i} = \sum_{i \ne j}\gamma_j \frac{x_i-x_j}{r_{ij}^2}.
\end{eqnarray}
Assuming that all vortices have the same conserved peak vorticity
$\pm\omega$, these equations are supplemented with the following
merging rules consistent with NS simulations
\cite{Carnevale1991,Weiss1993,Benzi1992} and experiments: when two
like-sign vortices of radius $a_1$ and $a_2$ reach a distance
$r_c\sim a_1+a_2$, they merge keeping the energy $E\sim n\omega^2
a^4$ constant, so that the resulting vortex has a radius $a$
satisfying, $a^4=a_1^4+a_2^4$. The average radius then scales as
$a\sim(n\omega^2/E)^{-1/4}\sim t^{\xi/4}\ll R$, where $R\sim
n^{-1/2}\sim t^{\xi/2}$ is the typical distance between vortices.
Hence, for large time for which $a\ll R$, we expect that the application of
the point-like vortex model to two-dimensional decaying turbulence becomes
\textit{more and more justified}.

The different scaling relations obtained in this work rely on the
assumption that the vortices are characterized by a single length
scale $a$. Hence, our results will be valid provided the vortex
radius distribution $p(r,t)$ is {integrable or} marginally
integrable for $r\to 0$, \textit{i.e.} $p(r,t)\ll r^{-\delta}$, with
$\delta\leq 1$. Otherwise, when the vortex size distribution is
polydisperse (for $\delta>1$), one must introduce a cut-off $r_<$
for small radii in order to ensure that $p(r,t)$ is normalizable.
Then, $r_<$ and $a$ will scale differently with time, {thereby}
introducing another relevant length scale apart from $a$ and $R$. In
the case $\delta=1$, $r_<$ and $a$ scale identically, up to
logarithmic corrections in the time $t$, and the present theory
still applies. Starting from an incoherent initial background with
an initial energy spectrum peaked at a certain scale $k_0=1/a_0$,
experiments and NS simulations show that the radius distribution
remains integrable and is very often bell-shaped (in the NS of
\cite{dritschel}, it is however claimed that $\delta=1$) and
scale-invariant, hence satisfying
\begin{equation}
p(r,t)= \frac{1}{a(t)}g\left(\frac{r}{a(t)}\right),
\end{equation}
where $g$ is the scaling distribution and the prefactor ensures the
normalization of $p$. Similarly, the peak vorticity
distribution is found numerically and experimentally to be bell-shaped, so
that only a single vorticity scale $\omega$ must be introduced.

From now on, we place ourselves in the framework of the point vortex
model described above, for which we present analytical and numerical
results. Hopefully, and as claimed in the two preceding paragraphs,
this model is faithful enough to the original problem to apply our
results to decaying  turbulence in actual fluids. Our first aim in
Section \ref{sec2turb} is to evaluate analytically and numerically
the time $\tau_m$ -- hence the exponents $\alpha$ and $\xi$ --
necessary to observe a merging in the context of a purely
\emph{two-body} mechanism. In Section \ref{sec3turb}, we will then
address the relevance, at least at small vortex density $na^2\ll 1$,
of \emph{three-body} mergings (typically, a ballistic dipole hitting
a single vortex), which will be shown to ultimately dominate. The
numerical results of this second study will be supplemented with a
fully consistent analytic mean-field three-body kinetic theory
showing that the asymptotic density decay exponent should be
$\xi=1$. We finally apply a similar analytic argument to
three-dimensional geostrophic decaying turbulence, finding an
exponent $\xi=5/4$ in excellent agreement with direct NS simulations
\cite{geo}. The present work hence precisely assesses the crucial
differences between a vortex dynamics dominated by two-body or
three-body (a dipole and an isolated vortex) merging processes,
demonstrating the leading role of the later at low vortex density
{(the importance of three-body interactions in sufficiently dilute
2D turbulence was mentioned in \cite{dritdev} and incorporated in
kinetic models of decaying 2D turbulence in
\cite{Sire2000,dritschel}, albeit in a different manner)}.

\section{Theory of two-body merging processes}\label{sec2turb}

We {first} consider the effective dynamics of two interacting
like-sign vortices, $i=1,2$, which are also advected by the other
vortices, assumed to be uniformly distributed in space but remaining
at a distance larger than $R$ from both tagged vortices (two-body
merging mechanism). Since we are ultimately interested in the regime
$r=|{\bf r}_1-{\bf r}_2|\sim a\ll R$, one can expand
Eqs.~(\ref{eqx},\ref{eqy}) in powers of the relative distance
between both vortices, $r$. After expressing the time in unit of
$\tau$ and distances in unit of $R\sim n^{-1/2}$, we find that the
relative distance between both vortices ${\bf r}={\bf r}_1-{\bf
r}_2$ obeys
\begin{eqnarray}
\label{stor}
\frac{d r}{d t} &=& r [\cos(2\varphi)\eta_\alpha +\sin(2\varphi) \eta_\beta],\\
\label{stop}
\frac{d \varphi}{d t} &=& \frac{1}{r^2} - \sin(2\varphi)
\eta_\alpha + \cos(2\varphi) \eta_\beta,
\end{eqnarray}
where we have used  cylindrical coordinates ${\bf
r}=(x,y)=(r,\varphi)$ and the fictitious particle at ${\bf r}$ is
now confined to a disk of radius $R=1$. Furthermore, $\eta_\alpha$ and
$\eta_\beta$ represent the effective noise due to all other
vortices, and are explicitly given by
\begin{equation}\label{etadef}
\eta_\alpha=  \sum_{j \ne 1,2} \gamma_j \frac{2 x_{0j}
y_{0j}}{r_{0j}^4}, \quad\eta_\beta= \sum_{j \ne 1,2} \gamma_j
\frac{y_{0j}^2 - x_{0j}^2}{r_{0j}^4},
\end{equation}
where ${\bf r}_0=(x_0,y_0)$ is the position of the center of mass of
the two vortices. In the absence of the other vortices
($\eta_\alpha=\eta_\beta=0$), the two tagged vortices would rotate
around each other remaining at a constant distance $r$. Moreover,
{even in the presence of the other vortices,} ${dr/dt}\to 0$ as
$r\to 0$, since if both vortices were at the same position, they
would experience the very same advection {from the other vortices}.
The noises can be shown to satisfy \cite{Sire2000}
\begin{equation}\label{eta2}
 \langle\eta_\alpha^2(t) \rangle = \langle\eta_\beta^2(t) \rangle\sim
 \tau^{-2}, \quad \langle\eta_\alpha(t)\eta_\beta(t')\rangle=0,
\end{equation}
where $\tau^{-2}\sim{\cal O}(1)$ in our dimensionless units.
Moreover, these noises only vary notably on the time scale over
which the other vortices travel a distance of order $R$. Their
correlation time is thus of order $\tau\sim{\cal O}(1)$
\cite{Sire2000}. The fact that this correlation time is non-zero (as
it would be for a standard pure white noise) is crucial: when the
two vortices are very close to each other, their mutual rotation
period will become much smaller than $\tau$, which will result in
the effective averaging out of the external noises
$\eta_{\alpha,\beta}$. Finally, it appears reasonable to modelize
$\eta_\alpha$ and $\eta_\beta$ by effective Ornstein-Uhlenbeck
processes \cite{Sire2000,McWilliams1998} with variance and
correlation time both equal to 1:
\begin{equation}\label{eta}
\frac{d \eta_\gamma}{d t} = - \eta_\gamma + \sqrt{2} w_\gamma,
\quad \gamma=\alpha,\beta,
\end{equation}
where $w_\alpha$ and $w_\beta$ are two independent
$\delta$-correlated Gaussian white noises, implying $
\langle\eta_\gamma(t)\eta_{\gamma'}(t')\rangle={\rm e}^{-|t-t'|}
\delta_{{\gamma},{\gamma'}}. $ The Fokker-Planck equation associated
to the system of Eqs.~(\ref{stor},\ref{stop},\ref{eta}), which
describes the evolution of the probability distribution function
(\textit{pdf}) of the variables
$(r,\varphi,\eta_\alpha,\eta_\beta)$,  can be easily written {(see
Appendix \ref{secBturb})}. Unfortunately, it seems unlikely  that
its general time-dependent solution can be analytically obtained.
However, it is straightforward to check that $
P(r,\varphi,\eta_\alpha,\eta_\beta)\propto r\, {\rm
e}^{-\eta_\alpha^2/2-\eta_\beta^2/2} $ is a stationary solution.
Hence, the relative coordinate ${\bf r}=(x,y)$ has a uniform
asymptotic distribution.

For $r\ll 1$, we can still obtain some useful information on
$P(r,t)$, since the stochastic term in Eq.~(\ref{stop}) can be
neglected with respect to $r^{-2}$ in this regime. Dividing
Eq.~(\ref{stor}) by Eq.~(\ref{stop}), and integrating over
$\varphi$, we obtain
\begin{equation}\label{an1}
\frac{1}{r^2(t)}-\frac{1}{r^2(0)}\sim\int_{\varphi_0}^{\varphi}
[\cos(2\varphi') \eta_\alpha(\varphi') + \sin(2\varphi') \eta_\beta(\varphi')]
\,d\varphi'.
\end{equation}
Hence, for small $r$, the variable $r^{-2}$ has approximately a
Gaussian distribution (being the linear sum of Gaussian variables
$\eta_\alpha$ and $\eta_\beta$) of variance
\begin{equation}\label{an2}
\int_{\varphi_0}^{\varphi}d\varphi_1\int_{\varphi_1}^{\varphi}d\varphi_2\,
\cos[2(\varphi_2-\varphi_1)]{\rm e}^{-|t(\varphi_2)-t(\varphi_1)|}.
\end{equation}
Since $t'(\varphi)\approx r^2(\varphi)\ll 1$, we can write
$t(\varphi_2)-t(\varphi_1)\sim (\varphi_2-\varphi_1)t'(\varphi_1)$.
Finally, after computing the integral in Eq.~(\ref{an2}) in the
limit of large $\phi$,
\begin{eqnarray}
\int_{\varphi_1}^{\varphi}d\varphi_2\,
\cos[2(\varphi_2-\varphi_1)]{\rm e}^{-(\varphi_2-\varphi_1)t'(\varphi_1)}&=&
\frac{t'(\varphi_1)}{4+t'^2(\varphi_1)}\nonumber\\
&\approx & \frac{t'(\varphi_1)}{4},
\end{eqnarray}
we find, using Eq.~(\ref{an1}), and for large $\varphi$ and time $t$
\begin{equation}\label{r0}
\left\langle\left(\frac{1}{r^2(t)}-\frac{1}{r^2(0)}
\right)^2\right\rangle\sim \frac{1}{4}\int_{\varphi_0}^{\varphi}t'(\varphi_1)d\varphi_1
\sim \frac{t}{4}.
\end{equation}
Remembering that the asymptotic \textit{pdf} is $P(r)\sim r$ and
that the random variable $r^{-2}$ is approximately Gaussian for
small $r$, the above argument suggests that $P(r,t)$ can be written
as
\begin{equation}\label{prt}
P(r,t)=r f[r/r_0(t)],~r_0(t)\sim t^{-1/4},~
f(x)\sim {\rm e}^{-x^{-4}},
\end{equation}
where the typical scale $r_0(t)\sim t^{-1/4}$ has been deduced from
Eq.~(\ref{r0}).

\begin{figure}[htbp]
\centerline{
\includegraphics[width=7.5cm,angle=0]{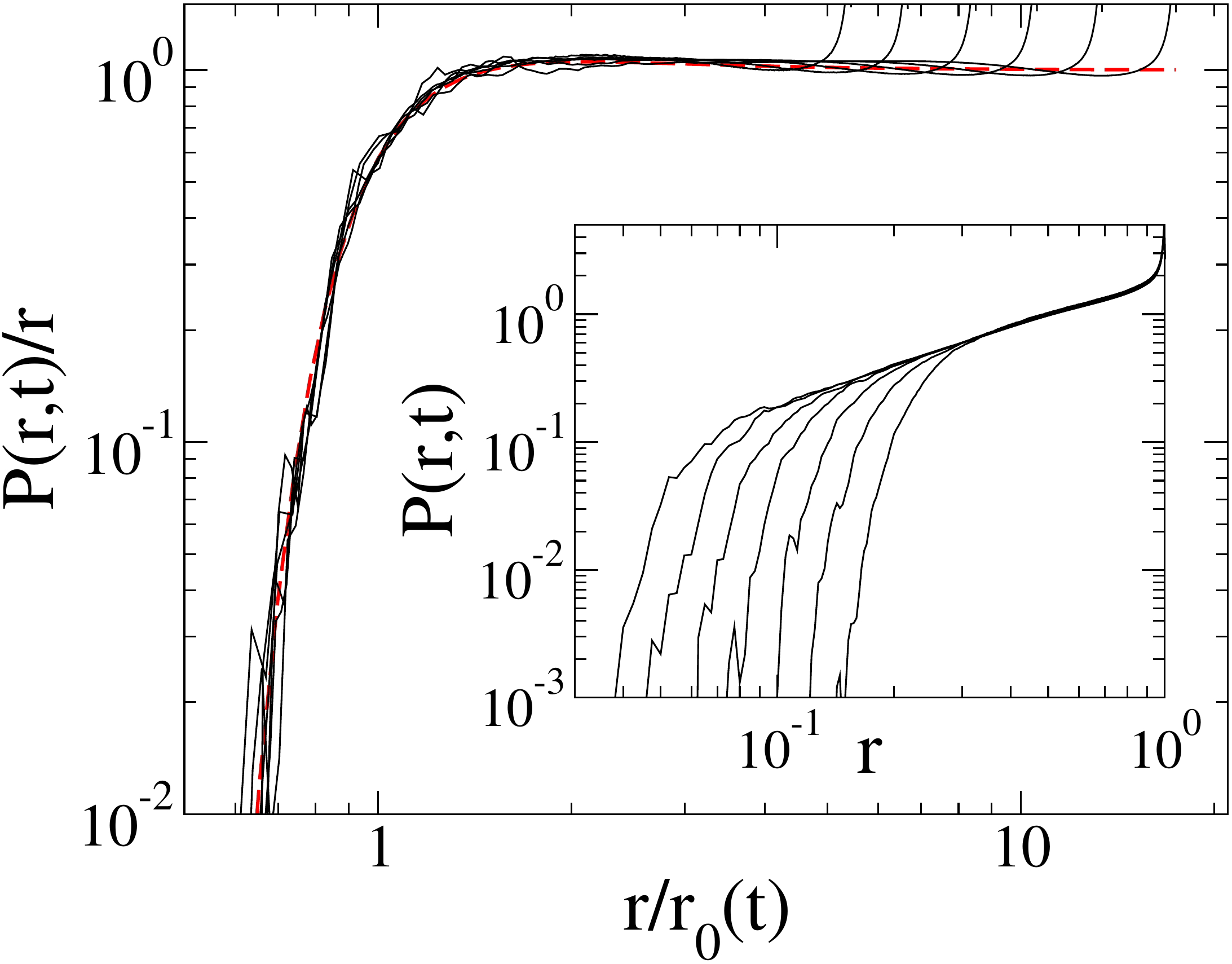}}
\caption[]{(Color online) Insert: we plot $P(r,t)/r$ for 7 different
times (in geometric progression) up to $t\sim 10^5$, averaged over
trajectories starting from
$r(0)\sim 1$. The main plot shows the corresponding data collapse,
fully consistent with Eq.~(\ref{prt}), along with a fit to the
predicted Gaussian functional form (in the variable $x^{-2}$) for the scaling function,
$f(x)={\rm e}^{-(x^{-2}-x_0^{-2})^2}$ ($x_0\approx 2.1$; dashed
line).} \label{fig_dens}
\end{figure}

This scaling behavior is illustrated in Fig.~\ref{fig_dens}, where
Eqs.~(\ref{stor},\ref{stop},\ref{eta}) have been solved numerically
using a second order stochastic integration scheme. We ran
$\sim$\,$10^5$ trajectories up to $t\sim 10^5$, starting from random
initial values for $r(0)\sim 1$ ($R=1$ in our units) and $\phi(0)$
(uniformly distributed in $[0,2\pi]$). The distribution of $r(t)$,
$P(r,t)$, is plotted for seven different times, illustrating the perfect
scaling collapse predicted by Eq.~(\ref{prt}).

\begin{figure}[htbp]
\centerline{
\includegraphics[width=7.5cm,angle=0]{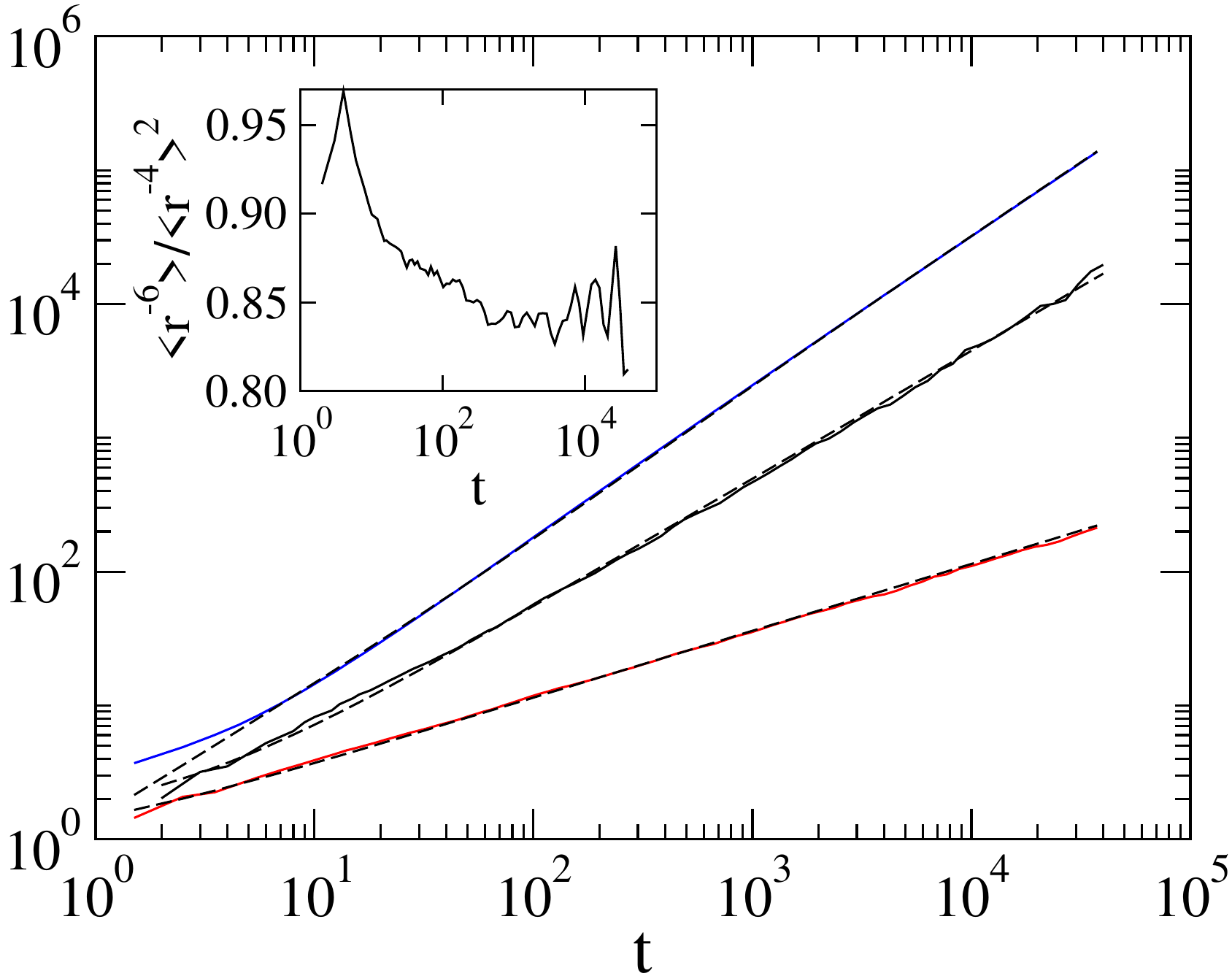}}
\caption[]{(Color online) From bottom to top, we plot
$m_{-4}(t)=\langle r^{-4}(t)\rangle$, $\exp(\langle
r^{-2}(t)\rangle)$ (which should have a power-law behavior, since
$m_{-2}(t)\sim \ln t$), and $\langle\varphi(t)\rangle$. The dashed
lines are fits to the predicted functional forms (see text):
$a_0t^{a_1}$ for $\exp[ m_{-2}(t)]$ and $m_{-4}(t)$ (in this
case, $a_1=0.49(1)$), and $a_0t\ln(t+a_1)$ for $\langle\varphi(t)\rangle$. The
Insert illustrates the near constancy of $m_{-6}(t)/m_{-4}^2(t)$,
which is also fully consistent with $r_0(t)\sim t^{-1/4}$ and the scaling
form of $P(r,t)$ in Eq.~(\ref{prt}).} \label{fig_mom}
\end{figure}

Using Eq.~(\ref{prt}), we can obtain the behavior of the moments of
$r(t)$ for large time. We find that for $z>-2$, $m_z(t)=\langle
r^{z}(t)\rangle$ converges to a constant, while $m_{-2}(t)\sim \ln
t$, and
\begin{equation}\label{momz}
m_z(t)\sim t^{-(z+2)/4},~{\rm for }~z<-2.
\end{equation}
In addition, Eq.~(\ref{stop}) implies $\langle\varphi(t)\rangle\sim
t\langle r^{-2}(t)\rangle\sim t\ln t$. These different predicted
behaviors are in excellent agreement with numerical simulations of
Eqs.~(\ref{stor},\ref{stop},\ref{eta}) presented in
Fig.~\ref{fig_mom}.

\begin{figure}[htbp]
\centerline{
\includegraphics[width=7.8cm,angle=0]{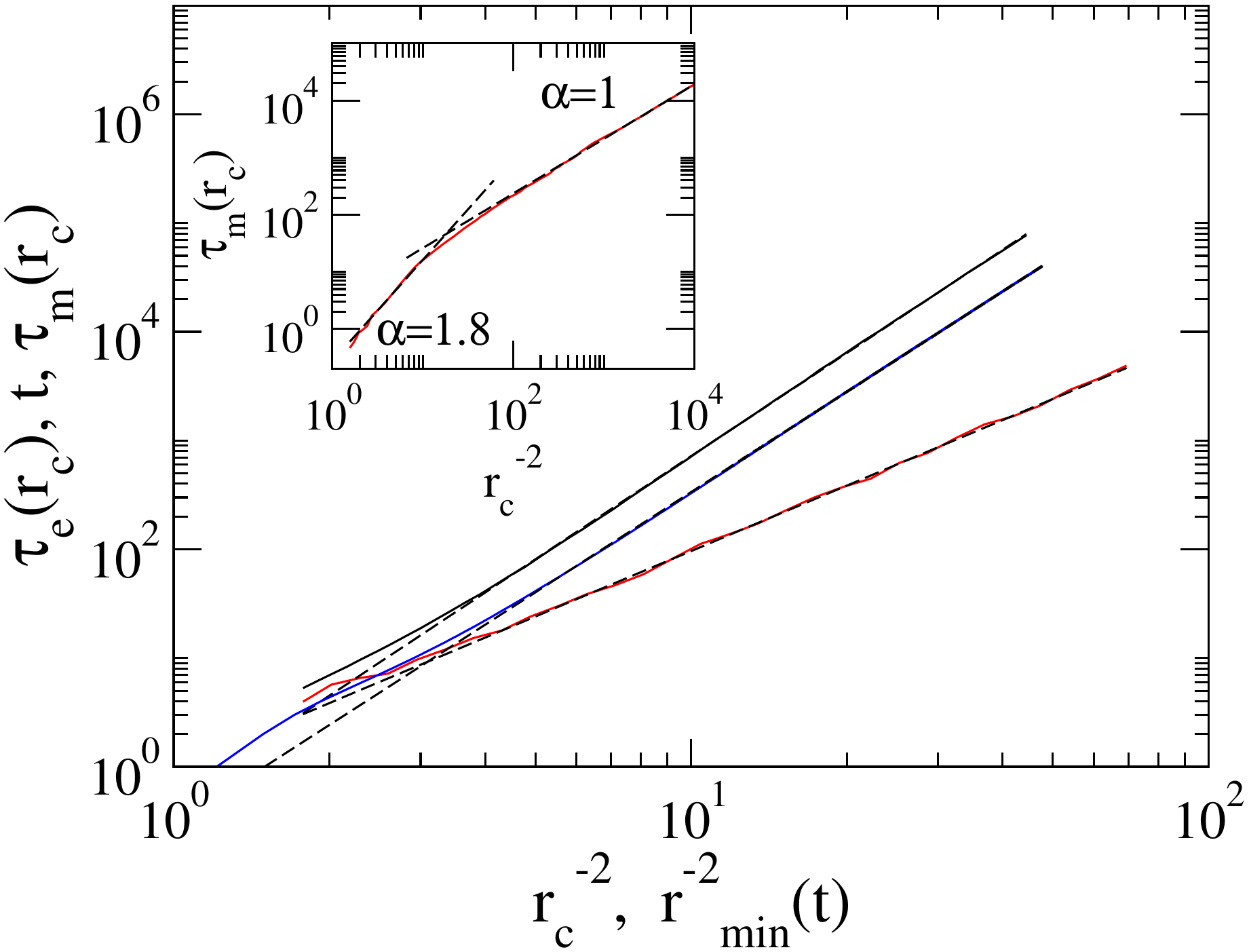}}
\caption[]{(Color online) For the
two-vortex effective model of Eqs.~(\ref{stor},\ref{stop},\ref{eta}),
we plot $\tau_m(r_c)$ \emph{vs} $r_c^{-2}$ (top full line) and the
time $t$ \emph{vs} $\langle r_{\rm min}^{-2}(t)\rangle$ (middle) along
with power-law fits with respective exponent $\alpha\approx 3.1$ and
$\alpha\approx 3.0$ (dashed lines). We also plot the escape time
$\tau_e(r_c)$ (bottom) and a fit using $\alpha=2$. These exponents are
in good agreement with the predictions of
Eqs.~(\ref{r0},{\ref{tescape}},\ref{tauexp}). For the
three-vortex model (Insert), we plot $\tau_m(r_c)$ and power-law fits
with exponent $\alpha\approx 1.8$ for ``large'' $r_c$ (high density)
and the predicted asymptotic exponent $\alpha= 1$, {for
$r_c\rightarrow 0$ (low density)}.}
\label{fig_tau}
\end{figure}

Let us now address the problem of determining the merging time
$\tau_m(r_c)$ for two vortices of radius $r_c$ starting at a
distance $r(0)\sim 1$, both lengths being expressed in unit of $R$. By
definition,
$\tau_m(r_c)$ is the average time necessary for the distance $r$ between
the two vortices to reach the value  $r_c$ for the
first time. Alternatively, defining $r_{\rm min}(t)$ as the minimum
$r(t)$ reached up to time $t$, one should have
\begin{equation}
\langle r_{\rm
min}(\tau_m(r_c))\rangle\sim r_c,
\end{equation}
and one can obtain an estimate of
$\tau_m(r_c)$ by inverting this relation. Moreover, our previous
heuristic argument resulting in Eq.~(\ref{r0}) actually shows that
the average escape time $\tau_e(r_c)$ to go from a region for which
$r(0)\in[(1-\varepsilon)r_c,r_c]$ (with $0<\varepsilon\ll 1$) to
another for which $r(t)\in[1-\varepsilon,1]$ is given by
\begin{equation}\label{tescape}
\tau_e(r_c)\sim r_c^{-4}.
\end{equation}
A (Markovian) detailed balance argument
then implies that the typical time to go from
$r(0)\in[1-\varepsilon,1]$ to $r(t)\in[(1-\varepsilon)r_c,r_c]$ -- a
time we assimilate to $\tau_m(r_c)$ -- satisfies
\begin{equation}\label{db}
\frac{{\rm Prob}(r\in[1-\varepsilon,1])}{\tau_m(r_c)}\sim
\frac{{\rm Prob}(r\in[ (1-\varepsilon)r_c,r_c])}{\tau_e(r_c)}.
\end{equation}
Using the stationary \textit{pdf}, $P(r)\sim r$, and the above result for
$\tau_e(r_c)$, we obtain
\begin{equation}\label{tauexp}
\tau_m(r_c)\sim r_c^{-6}, \quad\langle r_{\rm min}(t)\rangle\sim t^{-1/6}.
\end{equation}
We have numerically estimated $\langle r_{\rm min}(t)\rangle$ and
the first-passage times $\tau_m(r_c)$ and $\tau_e(r_c)$, by
integrating Eqs.~(\ref{stor},\ref{stop},\ref{eta}) using a second
order stochastic integration scheme (see Fig.~\ref{fig_tau}). $r_c$
being initially fixed, the merging time $\tau_m(r_c)$ is determined
by averaging over $\sim$\,$10^6$ (large $r_c$) to $\sim$\,$10^5$
(smallest $r_c$) {realizations}, the first time for which the
inter-vortex distance $r(t)<r_c$ (merging criterion), starting from
an initial $r(0)\sim 1$. Conversely, starting from an initial
$r(0)=r_c<r_{\rm max}$, the escape time $\tau_e(r_c)$ is estimated
as the first time for which $r(t)>r_{\rm max}\sim 1$ (we have set
$r_{\rm max} = 0.8$ in unit of $R$). The results presented in
Fig.~\ref{fig_tau} are in good agreement with our analytic estimates
of Eqs.~(\ref{tescape},\ref{tauexp}).

Reintroducing time and spatial units, we conclude that a purely
two-body mechanism would lead to $\alpha=3$ (see
Eqs.~(\ref{merg},\ref{tauexp})) and $\xi=1/2$. Even if we assimilate
the effective cut-off $r_0(t)\sim R(t/\tau)^{-1/4}$ to the vortex
radius $a(t)$, we end up with $\alpha=2$ and $\xi=2/3$. We will show
below that a three-body merging mechanism leads to a faster density
decay, and hence should dominate the dynamics at low vortex density.
Actually, the inefficiency of the two-body merging processes has a
simple physical interpretation suggested by our analytical results
of Eqs.~(\ref{an1},\ref{an2},\ref{r0}): when two like-sign vortices
are at a distance $r\sim a\ll R$, they rotate around each other at
such a high velocity (${d \varphi}/{d t}\sim r^{-2}$) that the
effective noise due to vortices at a distance greater than $R$
(which has a correlation time of order $\tau\sim 1$, in our units)
\textit{is averaged out by the fast rotation}. In other words, the
fast rotation makes the effective noise felt by both tagged vortices
so weak that the two-body system becomes almost integrable (in the
absence of noise, the distance between both vortices would remain
strictly constant). This points to the necessity of the presence of
at least another vortex at a distance of order $a$, in order to
significantly perturb the nearly integrable two-body system.

\section{The role of tree-body merging processes}\label{sec3turb}

By generalizing our model of Eqs.~(\ref{stor},\ref{stop},\ref{eta})
to the case of 3 vortices (2 like-sign vortices and 1 of opposite
sign), we define again $\tau_m(r_c)$ as the time for which the
distance between both like-sign vortices reaches the value $r_c$ for
the first time. We numerically obtain  that $\tau_m(r_c)\sim
r_c^{-2}$ (see below for a theoretical argument), leading to
$\alpha=1$ and $\xi=1$. The data are presented in the Insert of
Fig.~\ref{fig_tau}, confirming the results obtained in
\cite{Sire2000}, although for a wider range of values {of} $r_c$.
However, we find that the high density regime, corresponding to the
largest values of $r_c$, is best fitted by $\alpha\approx 1.8$ (see
Insert of Fig.~\ref{fig_tau}). This value of $\alpha$ is associated
to $\xi\approx 0.71$, in good agreement with (high density)
simulations and experiments. Moreover, we find that when the two
like-sign vortices are at a distance $r_c\ll 1$, the third
(opposite-sign) vortex is at a distance proportional to $r_c$
itself, justifying that this third vortex plays a crucial role in
the collision process. Hence, we have shown that considering
three-body merging processes speeds dramatically the dynamics, at
least at sufficiently low densities $na^2\ll 1$. {This leads to a
true scaling exponent $\xi=1$ (associated with $\alpha=1$).} This
regime is, {however,} difficult to reach in NS simulations and
experiments, and we claim that only a pseudo-scaling is observed at
the densities achieved, $na^2\sim 0.1$. This corresponds to a
transient regime with a pseudo-scaling exponent $\xi\simeq 0.71$
(associated with $\alpha\simeq 1.8$) measured on less that one
decade. This pseudo-scaling exponent is expected to slowly increase
with time, asymptotically reaching the true scaling exponent
$\xi=1$.  A strong evidence that true scaling is not reached in the
experiment of \cite{Hansen1998} is that the measured merging time
was found to scale as $\tau_m\sim t^{0.57\pm 0.12}$, instead of the
necessary $\tau_m\sim t$. In addition, an indication that the
exponent $\xi$ measured in direct numerical simulations slowly
increases with time is given in \cite{Laval2001}. In order to reach
the true scaling regime, one would need to run longer simulations
with more vortices at the start which is difficult. Alternatively, a
dynamical renormalization group analysis \cite{Sire2000} allows one
to reach very low densities more easily. In this case, it is found
that $\xi=1$ (measured on four decades) in agreement with the
present three-body kinetic theory.

Note that the local energy (see Eq.~(\ref{hamilt})) involving three
vortices of respective circulation $\gamma$, $\gamma$, $-\gamma$, is
${\cal H}_3\sim \gamma^2\ln(r_{13}r_{23}/r_{12})$. It can remain
finite and even constant even if $r_{12}\ll 1$, provided the third
vortex of opposite sign is in the vicinity of the two merging
vortices. In the case of a binary merging, the local energy ${\cal
H}_2\sim -\gamma^2\ln(r_{12})$ diverges when $r_{12}\ll 1$ and the
excess energy must be absorbed by the far away vortices, a very slow
process, as shown in the present work. A similar argument explains
why a three-vortex merging process involving three like-sign
vortices is as inefficient as a two-body process, as observed
numerically in \cite{Sire2000}. Finally, we have confirmed
numerically {(simulation not shown)} that adding a fourth vortex in
our effective model does not change the scaling of the Insert of
Fig.~\ref{fig_tau}, provided that all four vortices do not have the
same sign.

At low density $na^2\ll 1$, the proposed physical picture is that
ballistic dipoles collide with isolated vortices, the later
ultimately merging with the like-sign vortex of the dipole. We
briefly recall the kinetic theory presented by two of us in
\cite{Sire2000} and generalize it to any dimension $d$, and in
particular to geostrophic decaying turbulence of a three-dimensional
stratified fluid \cite{geo}, where planar vortices are replaced by
almost spherical ``blobs''. The vortex density satisfies an equation
of the form $dn/dt=-n/\tau_{m}$ and, as already mentioned, the
density power-law decay $n\sim t^{-\xi}$ implies $\tau_{m}\sim t$.
Noting that the dipoles move almost ballistically with velocity
$v_{dip}\sim \gamma/a\sim\omega a$ and using a classical cross
section argument, we obtain
\begin{equation}\label{taudip}
\tau_{m}\sim 1/(n_{dip} a^{d-1} v_{dip}),
\end{equation}
where $d$ is the dimension of space. Using a mean-field
approximation, dipoles of typical size $a$ are formed with a density
$n_{dip}\sim n{\times} na^d$. Moreover, the energy $E\sim (\omega a)^2 n
a^d$ and the (potential) vorticity peak $\omega$ are conserved, so
that Eq.~(\ref{taudip}) leads to
\begin{equation}\label{taudipd}
\tau_{m}\sim \omega^{-1}\left(\frac{\omega^2}{E}\right)^{\frac{2d}{d+2}}
{\times} n^{-\frac{4}{d+2}},
\end{equation}
where we have kept track of the dimensional constants $\omega$ and
$E$. Equivalently, this result can be rewritten in the same form as
in Eq.~(\ref{merg})
\begin{equation}\label{taudipda}
\tau_{m} \sim \frac{\tau}{\left(n a^2\right)^\frac{6-d}{d+2}}
\end{equation}
Finally, expressing the asymptotic scaling condition, $\tau_{m}\sim
t$, Eqs.~(\ref{taudipd},\ref{taudipda}) lead to
\begin{equation}
\xi=\frac{d+2}{4},\quad \alpha=\frac{6-d}{d+2}.
\end{equation}

In $d=2$, we obtain $\xi=1$ (associated to $\alpha=1$), which agrees
with the numerical results of Fig.~\ref{fig_tau}. Again, this value
is slightly higher than the exponents measured in experiments and
most numerical simulations. However, we have already argued that the
smallest area fraction $na^2$ reached remains quite large, except in
the RG procedure of \cite{Sire2000}, which indeed found $\xi\approx
1$. In the experiment of \cite{Hansen1998}, the measured merging
time was found to scale as $\tau_m\sim t^{0.57\pm 0.12}$, instead of
the necessary $\tau_m\sim t$, showing that the scaling regime was
not yet attained. However, we observe that the theoretical relation
of Eq.~(\ref{taudipd}) is remarkably well verified in this
experiment:
\begin{equation}
\tau_m\sim t^{0.57\pm 0.12}\sim \frac{\omega}{E}{\times} n^{-1}\sim t^{0.55\pm 0.14}.
\end{equation}
This suggests that the result of Eq.~(\ref{taudipd}) could be valid
in a wider range of time, even before reaching the true asymptotic
scaling regime. This phenomenon is actually quite common in the
context of critical phenomena, when dealing with finite-size samples
and simulation time: the scaling relations between critical physical
quantities (like the susceptibility or the correlation length/time
in magnetic systems) are often found to be better obeyed numerically
than between these quantities and the critical parameter itself
(\emph{e.g.} the distance to the critical temperature $|T-T_c|$).

In $d=3$, we obtain $\xi=5/4$ (associated to $\alpha=3/5$), in good
agreement with the value $\xi\simeq 1.25\pm 0.10$ measured in NS
simulations of geostrophic decaying turbulence in a stratified fluid
\cite{geo}. We argue that due to the fastest decay in $d=3$, the low
density regime where the three-body mechanism sets in appears
faster. In addition, the increased dimensionality generally plays in
favor of a better validity of a mean-field treatment.

\section{Conclusion}\label{sec4turb}

In conclusion, we have presented an effective theory for the merging
of two vortices advected by the effective noise of the other
vortices. Several quantities including the merging time $\tau_m$ and
the \textit{pdf} $P(r,t)$ have been derived analytically and are in
excellent agreement with numerical simulations. The results of this
study point to the weak efficiency of two-body processes, as the
fast rotation of the vortex pair averages out the effective
advecting noise due to the other vortices. We numerically find that
a similar effective theory, where three vortices {of different sign}
now participate to the merging process, leads to a faster decay of
the density, $n(t)\sim t^{-1}$, after a transient regime where
$n(t)\sim t^{-0.71}$. We {therefore} conclude that three-body
processes should dominate, at least at small vortex density. This
result can be understood within a simple mean-field kinetic theory
describing the merging process as the collision of a ballistic
dipole with an isolated vortex. This same theory applied to
geostrophic turbulence in three dimensions leads to $n(t)\sim
t^{-5/4}$, in good agreement with numerical simulations. It would be
certainly interesting to generalize this study for an increasing
number $N>4$ of tagged vortices, yet small enough for the numerical
simulations to reach very small effective densities (\emph{i.e.}
small $r_c$). The $N$ vortex mutual dynamics would still be treated
exactly, but in the effective noisy background generated by all the
other vortices.

\appendix

\section{Conservation of the vorticity peak}\label{secAturb}

It is numerically observed that the vorticity peak is almost
conserved in 2D decaying turbulence \cite{Carnevale1991}, an
important ingredient of the scaling theory. In this Appendix, we
would like to motivate this claim in terms of statistical mechanics
and kinetic theory of the 2D Euler equation.

The Miller-Robert-Sommeria (MRS) statistical theory \cite{miller,rs}
is relatively well-suited to describe the rapid merging of two
vortices. Knowing the initial condition and assuming ergodicity
(efficient mixing) the statistical theory predicts the shape of the
new vortex resulting from this merging. Furthermore, the mixing
process can be described in terms of relaxation equations
\cite{rsmepp} obtained from a maximum entropy production principle
(MEPP). Assuming, for simplicity, that the initial vortices have a
uniform vorticity $\omega=\sigma_0$, the relaxation equations can be
written
\begin{equation}
\label{w1}
\frac{\partial\overline{\omega}}{\partial t}+{\bf u}\cdot
 \nabla\overline{\omega}=\nabla\cdot \lbrack D({\bf r},t)
 \left (\nabla\overline{\omega}+\beta(t)\overline{\omega}
 (\sigma_0-\overline{\omega})\nabla\psi\right )\rbrack,
\end{equation}
\begin{equation}
\label{w2}
\beta(t)=-\frac{\int D({\bf r},t)\nabla\overline{\omega}\cdot\nabla\psi\,
d{\bf r}}{\int D({\bf r},t)\overline{\omega}(\sigma_0-\overline{\omega})
(\nabla\psi)^2\, d{\bf r}},
\end{equation}
where $\overline{\omega}({\bf r},t)$ is the coarse-grained vorticity
field and $D({\bf r},t)$ the ``turbulent'' diffusion coefficient. If
$D({\bf r},t)$ is constant, the relaxation equations converge
towards the statistical equilibrium state given by a
Fermi-Dirac-like distribution
\begin{equation}
\label{w3}
\overline{\omega}({\bf r})=\frac{\sigma_0}{1+e^{\sigma_0(\beta\psi({\bf r})+\alpha)}}.
\end{equation}
However, it has been observed at several occasions (see, e.g.,
\cite{chen,brands}) that relaxation is {\it incomplete} due to lack of
ergodicity. This is clearly a limitation of the statistical theory. In
order to take incomplete relaxation into account, it has been proposed
that the diffusion coefficient $D({\bf r},t)$ varies in space and
time. Qualitative arguments \cite{rr}, or more sophisticated kinetic
theory \cite{quasi}, indicate that the diffusion coefficient should be
proportional to the local variance
$\omega_2=\overline{\omega^2}-\overline{\omega}^2=
\overline{\omega}(\sigma_0-\overline{\omega})$
of the vorticity fluctuations. It must also decrease in time as the
fluctuations decay. Therefore
\begin{equation}
\label{w4}
D({\bf r},t)=A(t) \overline{\omega}({\bf r},t)\lbrack
\sigma_0-\overline{\omega}({\bf r},t)\rbrack,
\end{equation}
where $A(t)$ tends to zero on a few dynamical times. In the ``mixing
zone'' where the fluctuations are strong, the diffusion coefficient
is large and the system rapidly reaches a Fermi-Dirac-like
distribution (\ref{w3}), with an inverse temperature $\beta'$ and a
chemical potential $\alpha'$ that may be different from the
equilibrium ones. On the other hand, from Eq. (\ref{w4}), one sees
that the diffusion coefficient is small in the regions where the
fluctuations are weak. This concerns the core of the vortex
($\overline{\omega}({\bf r},t)\rightarrow\sigma_0$) where the
vorticity is maximum and the region surrounding the vortex
($\overline{\omega}({\bf r},t)\rightarrow 0$) which is not sampled
by vorticity. Therefore, the relaxation towards the Fermi-Dirac-like
distribution (\ref{w3}) will be very slow in these regions. Since,
in parallel, the diffusion coefficient decreases with time (because
of the term $A(t)$), the system will not have time to reach the
Fermi-Dirac-like distribution in these regions. Therefore, mixing
will be relatively inefficient in the core and at the periphery of
the vortex. These arguments of kinetic theory may explain why the
vorticity peak is relatively well conserved during a merging and why
the vortex is more confined than what the statistical theory
predicts assuming {\it complete} mixing. These arguments have been
developed in Sec. 4.3 of \cite{bbgky}.

\section{The Fokker-Planck equation}\label{secBturb}

In order to write down the Fokker-Planck equation associated with
the  stochastic process defined by
Eqs.~(\ref{stor},\ref{stop},\ref{eta}), it may be simpler to use
cartesian coordinates. Equations (\ref{stor},\ref{stop}) are
therefore rewritten as
\begin{equation}
\frac{dx}{dt}=-\frac{y}{r^2}+x\eta_\alpha+y\eta_\beta,
\end{equation}
\begin{equation}
\frac{dy}{dt}=\frac{x}{r^2}-y\eta_\alpha+x\eta_\beta.
\end{equation}
Introducing the convenient notations  ${\bf r}=(x,y)$ and ${\bf
w}=(\eta_\alpha,\eta_\beta)$, the Fokker-Planck equation for $P({\bf
r},{\bf w},t)$ is
\begin{eqnarray}
\frac{\partial P}{\partial t}+\frac{\partial}{\partial x}\left\lbrack
\left (-\frac{y}{r^2}+x \eta_\alpha+y\eta_\beta\right )P\right\rbrack+\nonumber\\
\frac{\partial}{\partial y}\left\lbrack \left (\frac{x}{r^2}-y
\eta_\alpha+x\eta_\beta\right )P\right\rbrack=\frac{\partial}{\partial {\bf w}}
\cdot\left (\frac{\partial P}{\partial {\bf w}}+P{\bf w}\right ).\quad
\end{eqnarray}
The stationary solution ($\partial_t P=0$) of  this Fokker-Planck
equation is obtained when the r.h.s. and the l.h.s. vanish
individually. The cancelation of the r.h.s. implies that
\begin{equation}
P_s({\bf r},{\bf w})=C({\bf r})e^{-w^2/2}.
\end{equation}
Inserting this expression in the l.h.s. of the Fokker-Planck
equation,  we obtain $C({\bf r})={\rm constant}$. Therefore,
\begin{equation}
P_s({\bf r},{\bf w})\propto e^{-w^2/2}.
\end{equation}
It is noteworthy that the steady state of this Fokker-Planck
equation is spatially homogeneous, \emph{i.e.} $P_s({\bf r})={\rm
constant}$, and $P_s(r)=r$ in radial coordinates.

\end{document}